\documentclass[reprint,amsmath,amssymb,aps,pra]{revtex4-1}

\usepackage{graphicx}
\usepackage{dcolumn}
\usepackage{bm}

\begin{document}

\title{Resonant spin wave excitation in magnetoplasmonic bilayers by short laser pulses}

\author{Stanislav Kolodny}
\affiliation{ITMO University, Saint Petersburg 197101, Russia}
\author{Dmitry Yudin}
\affiliation{ITMO University, Saint Petersburg 197101, Russia}
\author{Ivan Iorsh}
\affiliation{ITMO University, Saint Petersburg 197101, Russia}

\date{\today}

\begin{abstract}
In magnetically ordered solids a static magnetic field can be generated by virtue of the transverse magneto-optical Kerr effect (TMOKE). Moreover, the latter was shown to be dramatically enhanced due to the optical excitation of surface plasmons in nanostructures with relatively small optical losses. In this paper we suggest a new method of resonant optical excitations in a prototypical bilayer composed of noble metal (Au) with grating and a ferromagnet thin film of yttrium iron garnet (YIG) via frequency comb. Based on magnetization dynamics simulations we show that for the frequency comb with the parameters, chosen in resonant with spin-wave excitations of YIG, TMOKE is drastically enhanced, hinting towards possible technological applications in the optical control of spintronics systems. 
\end{abstract}

\maketitle

\section{Introduction}
Ideas and concepts developed in photonics have been recently applied in related fields, in particular in physics of collinear magnets. The rapidly advancing area of magnonics represents photonics with spin waves, or, collective propagating magnetization excitations in magnetic materials. Thanks to their unique linear and nonlinear properties~\cite{Demokritov2012,Khitun2010,Haldar2017,Fischer2017,Tabuchi2015} spin waves can be successfully employed in next generation spintronics devices and quantum computing. In the meantime, further implication in technology requires excitation of spin waves at short length- and time-scales. This can be achieved by the optical excitation of magnetic system with femtosecond laser pulses~\cite{VanKampen2002,kimel2005ultrafast,Hansteen2006,Stanciu2007,Bigot2009,Satoh2012,Au2013a}. Despite a variety of ways of optical generation of spin waves, the nonthermal magneto-optical processes such as inverse Faraday and inverse Kerr are of particular importance, since they are characterized by high time resolution and do not require heating of the sample. The essence of the effect is that the external optical electric field $\bm{E}$ induces magnetic field $\bm{h}\propto\bm{E}^*\times\bm{E}$ via the nonlinear processes such as e.g. stimulated Raman scattering~\cite{Popova2012}. Thus induced magnetic field drives spin waves in a magnetic structure. Recently, it was anticipated that in inhomogeneous media the expression $\bm{E}^*\times\bm{E}$ may be non-zero even for the linearly polarized wave. In planar geometry, this effect takes place for TM polarization only, and the induced magnetic field is perpendicular to the plane of incidence, realizing thus the inverse \textit{transverse} magneto-optical Kerr effect (TMOKE)~\cite{PhysRevB.86.155133}. Among structures which allow inverse TMOKE are magnetoplasmonic systems~\cite{Bossini2016}, since they facilitate electric field localization and an effective chirality $\sigma\propto\vert\bm{E}^*\times \bm{E}\vert/|\bm{E}|^2 $ of surface plasmon polariton can approach a unity. 

It is clear that in continuous wave regime the induced magnetic field $\bm{h}$ is time-independent, whereas under femtosecond laser pulses this field follows the intensity profile. Therefore, a much higher efficiency of spin wave generation can be achieved in the case of resonant excitation, when the intensity profile is periodic with a frequency close to that of ferromagnetic resonance, which is typically of the order of 1 GHz. Recently, it has been demonstrated experimentally~\cite{Jaeckl2017} that using clocked laser excitation can substantially enhance the efficiency of the spin wave generation at the bottom of magnon band, followed by their diffusion. However, for the direct implication in data processing resonant excitation of spin waves with fixed both frequency and wavevector (and group velocity, as a result) is of vital importance.

In this paper we propose a combination of a magnetoplasmonic structure and clocked laser excitation for the resonant excitation of spin waves. We first design the structure and numerically model the distribution of electromagnetic field under the clocked laser excitation. Using the data we evaluate the induced magnetic field and study magnetization dynamics based on numerical solution of Landau-Lifshitz-Gilbert (LLG) equation \cite{stancil2009spin}. We provide an estimate for a realistic structure, where a resonant excitation of spin waves can be achieved with efficiency enhancement of at least two orders of magnitude as compared to that in the free magnetic film under pulsed excitation.

\section{Model system}

Among various magnetic materials of collinear ordering yttrium iron garnet (YIG) is characterized by the best magneto-elecctrical and magneto-optical properties, which stand out this material for microwave applications \cite{Musa2007}. YIG belongs to the class of magnetically ordered structures with cubic symmetry \cite{pirro2014spin,collet2017spin} and rather low loss \cite{Pardavi2000}, and is successfully used in a broad range of spintronics applications. 
\begin{figure*}[ht]
\center{\includegraphics[width=\linewidth]{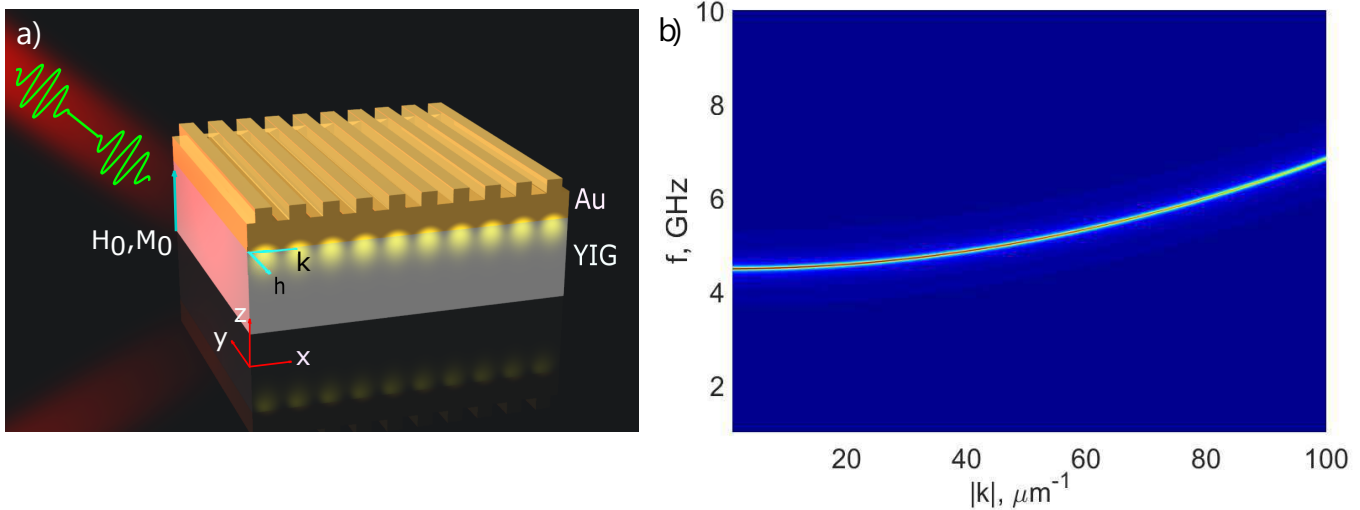}}
\caption{(a) A schematic representation of the bilayer under consideration: a thin film of ferromagnetic YIG with the magnetization $\bm{M}_0$ aligned along the $z$ axis is covered by a layer of gold with grating. The system is placed to the external magnetic field $\bm{H}_0$ directed normal to the interface; and is irradiated with the laser pulses of intensity $I$ and duration $t$, so that the induced magnetic field $\bm{h}$ in YIG is perpendicular to the plain of incidence ($xOz$). The latter is associated with surface plasmon polaritons emerging right at the interface between two metals. (b) The magnon dispersion relation in YIG obtained by numerical solution of Landau-Lifshitz-Gilbert equation.}\label{pic1}
\end{figure*}
We therefore consider a hybrid nanostructure consisting of a thin film of YIG and Au discrete grating on the top (geometry of the system is schematically depicted in Fig.~\ref{pic1}a under short laser pulses.

To study magnetization dynamics induced by laser pulses we are to numerically solve LLG equation 
\begin{equation}\label{llg}
\frac{\partial\bm{M}}{\partial t}=-\gamma\bm{M}\times\bm{H}_\textrm{eff}+\alpha\bm{M}\times\frac{\partial\bm{M}}{\partial t},
\end{equation} 
where $\gamma$ is the gyromagnetic ratio of an electron ($\gamma=28$ GHz/T). In general LLG equation describes precession of the magnetization, $\bm{M}$, around the effective magnetic field, $-\gamma\bm{H}_\textrm{eff}=\partial\mathcal{H}/\partial\bm{M}$, created by $d$ electrons of a ferromagnet, whereas the relaxation rate to the direction of the field is associated with the Gilbert damping parameter, $\alpha$. We assume the thin layer of YIG is initially polarized along the $z$ axis $\bm{M}_0=M_0\bm{e}_z$, as shown in Fig.~\ref{pic1}a, and is placed to the external field $\bm{H}_0$. Being irradiated with laser pulses results in the emergence of the magnetic field $\bm{h}$ directed perpendicular to the plain of incidence via TMOKE mechanism. The latter gives rise for the magnetization direction to slightly deviate from collinear ordering, and thus to spin waves excitation. To get spin wave spectrum we linearize Eq.~(\ref{llg}) with respect to in-plane components of the magnetization $\bm{m}=(m_x,m_y)$, i.e., we plug $\bm{M}=M_0\bm{e}_z+m_x\bm{e}_x+m_y\bm{e}_y$ into (\ref{llg}) on condition that $\vert m_x\vert,\vert m_y\vert\ll M_0$. We derive after straight forward algebra,
\begin{equation}
\frac{\partial\bm{m}}{\partial t}=-\gamma\bm{M}_0\times\bm{H}_\textrm{eff}+\frac{\alpha}{M_0}\left(\bm{M}_0\times\frac{\partial\bm{m}}{\partial t}\right). \label{eq1}
\end{equation}
The properties of YIG can be well approximated using the microscopic Hamiltonian which includes Heisenberg exchange interaction, magnetocrystalline anisotropy, and Zeeman coupling. Thus, the effective field can be represented as follows:
\begin{equation}
-\gamma\bm{H}_\textrm{eff}=\bm{h}+\bm{H}_0+\bm{H}_k+\alpha_{ik}\frac{\partial^2\bm{m}}{\partial x_i\partial x_k}-\hat{N}\bm{m},
\label{eq2}
\end{equation}
where we assume the summation over repeated indexes; $\hat{N}$ is the tensor of magnetocrystalline anisotropy, $\hat{N}\bm{m}=N_{ij}m_j$. In Eq.~(\ref{eq2}) the exchange tensor $\alpha_{ik}=A\delta_{ik}$ is reduced to a scalar owing to the cubic lattice symmetry of YIG with $A=$3 $\cdot$ 10$^{-16}$ m$^2$, while $\bm{H}_k$ corresponds to the effective anisotropy magnetic field \cite{stancil2009spin}. Typically, the effective anisotropy field $\bm{H}_k$ is material-specific and equals 4.6 kA/m for YIG. In the following, we assume $\bm{M}_0$ and $\bm{H}_k$ are parallel to $\bm{H}_0$ (see Fig.\ref{pic1}a), thus being normal to Au--YIG interface. We put $H_0=1$ T, and set $M_0$ to be equal to the saturation magnetization of YIG, i.e., $M_s=140$ kA/m. It is therefore clear that with no $\bm{h}$ present in the system the vector product $\bm{M}_0\times\bm{H}_{eff}$ is zero, producing no magnetization precession. 

\section{Results and discussion}

We put the Gilbert damping parameter $\alpha=$3.2 $\cdot$ 10$^{-4}$, although it can be decreased as shown in Ref.~\cite{dubs2017sub}. Fourier transforming the equation of motion (\ref{eq1}) to $\bm{m}(\omega,k)=G(\omega,k)\bm{h}(\omega,k)$ makes it possible to evaluate the spin wave spectrum from $G^{-1}(\omega,k)=0$. Thus obtained the magnon dispersion relation for the YIG thin film in the presence of $\bm{h}$ is shown in Fig.~\ref{pic1}b. One can clearly observe that magnons are positioned in 4-7 GHz range for $k$-vector ranging between 1 and 100 $\mu m^{-1}$), and are characterized by a very narrow dispersion line (about 40 MHz) because of low Gilbert damping. Therefore, the direct optical excitation of magnons is hardly to be achieved. 
\begin{figure*}[ht]
\center{\includegraphics[width=\linewidth]{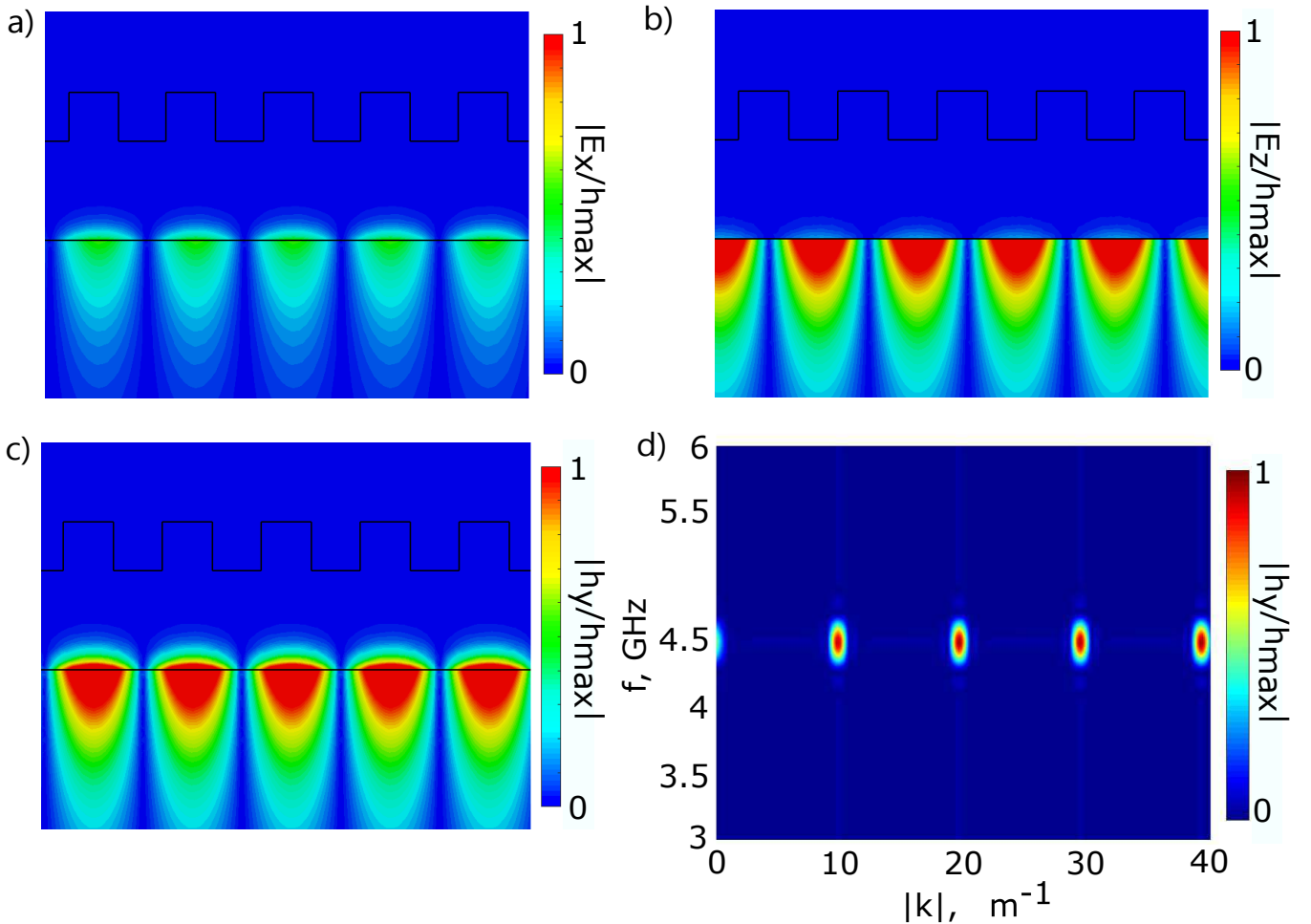}}
\caption{\label{pic2}Distribution of electric field components $E_x$ (a) and $E_z$ (b) inside the magnetoplasmonic bilayer, normalized by the maximum value of electric field. The induced magnetic field $\bm{h}\propto\bm{E}^\ast\times\bm{E}$ is strongly localized at the Au-YIG interface (c), normalized by the maximum value of the field. Dispersion of $\bm{h}$, created by the train of femtosecond optical pulses with repetition rate 4.5 GHz (d). Noteworthy, in these calculations we impose the periodic boundary conditions along $x$ and $y$ axes.}
\end{figure*}

\begin{figure*}[ht]
\center{\includegraphics[width=0.65\linewidth]{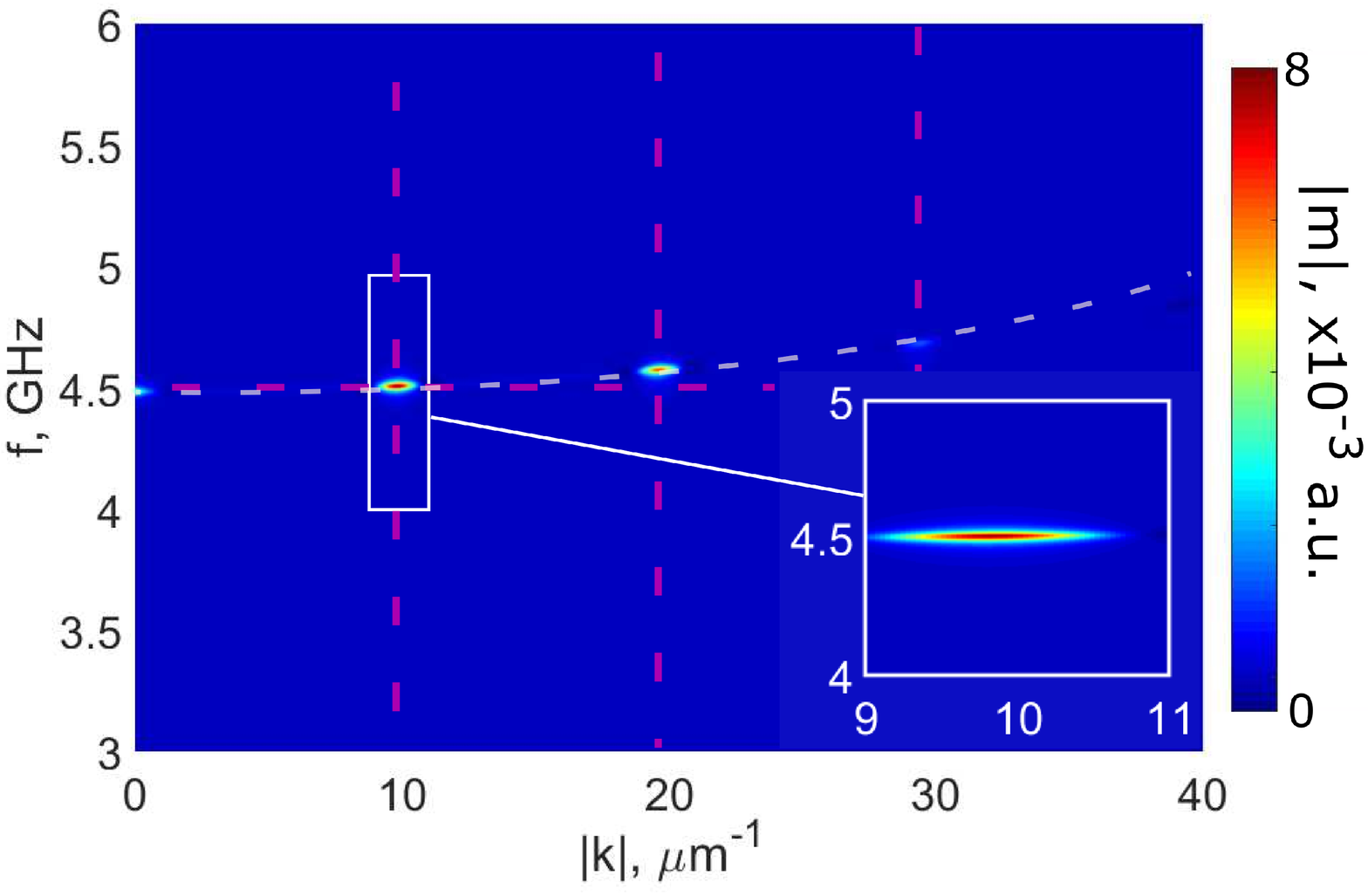}}
\caption{Magnon dispersion relation in hybrid Au-YIG nanostructure. White dashed curve represents the dispersion line of spin waves in YIG, crossing points of dashed lines represents the dispersion of $\bm{h}$.\label{pic3}}
\end{figure*}

To bind optical waves (in the form of plasmons, for example) with spin waves we propose to use a gold grating that amplifies the electric field and are to determine the field distribution due to strong field localization via excitation of plasmonic resonances \cite{maier2007plasmonics,giannini2011plasmonic}. To be more specific, we suppose: the thickness of Au layer $d$ is 100 nm, periodicity $a$ is 100 nm, the width $w$ and the height $h$ of array's element are 50 nm. The latter allows us to excite plasmons on the Au-YIG interface with the $k$-vector to be $10^7$ m$^{-1}$.To estimate the distribution of magnetic field inside the hybrid nanostructure we impose the periodic boundaries in $x$ and $y$ directions. We further proceed with eigenmode simulation in CST Microwave Studio: the electric field distribution clearly reveals that excited field of plasmons possesses two phase-shifted components of the electric field (see \ref{pic2}a-b) inside the YIG film. Thus, it induces time-independent magnetic polarization via the TMOKE mechanism $\bm{h}\propto\bm{E}^\ast\times\bm{E}$ \cite{PhysRevB.86.155133}.

The time-independent magnetic field $\bm{h}$ serves as a source of spin waves in accordance with Eq.~(\ref{eq2}). To introduce time-dependence we propose to excite plasmons in the nanostructure by a train of Gaussian pulses with the repetition rate chosen in resonance with the frequency of magnons in the YIG thin film. We assume the duration of each pulse is 40 ps that is shorter in comparison with the repetition rate. As we discussed earlier, the Au grating allows us to generate electromagnetic field with a fixed value of the wavevector $k=10^{7}$ m$^{-1}$, thus, in full accordance with the magnon dispersion of YIG shown in Fig.~\ref{pic1}a the resonance frequency of the spin waves in the magnetoplasmonic bilayer under consideration is 4.5 GHz. To be more realistic we put the finite size of hybrid nanostructure limited by 1 $\mu m $. Such a value of repetition rate is difficult to achieve using only common methods of Gaussian pulses train generation. Meanwhile, using a frequency comb technique facilitates overcoming this problem by generating the Gaussian pulses with high repetition rate up to THz range \cite{Wu:13,Kippenberg2011555}. The dispersion of induced magnetic field consists of a discrete set of spots as shown in Fig.\ref{pic2}d, the latter happens due to the periodicity of the structure and an excitation properties of frequency combs. Therefore, by solving the equation of motion with thus obtained $\bm{h}$ we can get the spin waves dispersion $\bm{m}(\omega,k)$ as well as to recover their time and space dependence by performing inverse discrete Fourier transformation. The resulting dispersion is depicted in Fig.~\ref{pic3}. As it was expected there is only one spot on coincidence of dispersion diagrams of excitation magnetic field and spin waves in YIG (hot spots of magnetic field diagram are represented by crossing of straight dashed pink lines and dispersion line of magnons in YIG is represented by white dashed curve).
\begin{figure}[ht]
\center{\includegraphics[width=\linewidth]{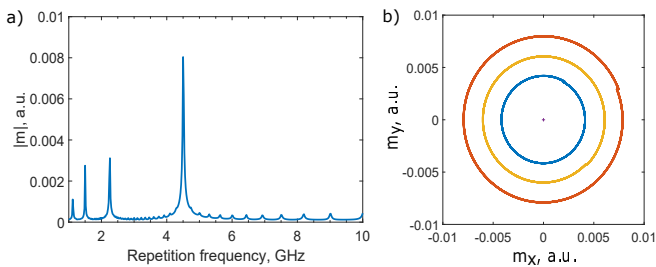}}
\caption{(a) Spin wave magnitude in the YIG thin film vs. repetition rate of femstosecond laser pulses. (b) Precession of $\vec{m}$ around equilibrium point in time for different excitation frequency: brown curve -- resonant excitation frequency 4.5 Ghz, blue curve -- 4.48GHz, yellow curve -- 4.52 GHz.}\label{pic4}
\end{figure}
The dependence of $\vert\bm{m}\bm$ on the repetition rate of femtosecond laser pulses for fixed values of $k$ in range 1-10 GHz for maximum value of excitation magnetic field $h/h_{max}$ is presented in Fig.~\ref{pic4}a, which clearly manifests the formation of a resonance mode. The resonance mode appears at 4.5 GHz. Also there are additional equidistant peaks are placed on lower frequencies. The main and additional peaks are very narrow because of low damping of spin waves in YIG which is described by Gilbert damping parameter as it was mentioned before. In addition, we study hodograph of the magnetization at different frequencies: In particular, results for 50 Gaussian pulses are shown in Fig.~\ref{pic4}b.
\section{Conclusions}
In the current studies we considered a typical magnetoplasmonic structure, that represents the bilayer of a thin film YIG covered by a layer of gold with grating and explicitly showed that TMOKE emerging in such a structure may be dramatically enhanced using frequency comb technique. In fact, for the repetition rate chosen to be in resonant with the ferromagnetic resonance of the magnetic layer the direct numerical solution to LLG equation reveals a desired behavior. We believe that our results will trigger experimental activity in rapidly advancing area of research right in the border between plasmonics and spintronics.
\section*{Acknowledgements}
We acknowledge the support from the Russian Science Foundation under the Project No. 17-12-01359.

\end{document}